# Phase Sensitive Amplifier Based on Ultrashort Pump Pulses

**Alexander Gershikov and Gad Eisenstein**

*Department of Electrical Engineering, Technion, Haifa, 32000, Israel.*
*Corresponding author: alexger@campus.technion.ac.il Tel. +97248294717*

**Abstract:** We demonstrate a narrow band phase sensitive amplifier in the pump degenerate configuration which employs ps pump pulses. Control of the amplifier bandwidth is achieved via changes of the pump spectral width. A phase sensitive gain between -6 and 6 decibels, with an overall system gain of 28dB was demonstrated.

## 1. Introduction

Parametric fiber amplifiers [1] are key element in optical networks as they offer high efficiency [2], robustness [3] and flexibililty [4]. Conventional, phase insensitive amplifiers (PIA) have a fundamental noise limitation of 3dB [5], however, operation in a phase sensitive mode [6] allows to achieve noise figures of less than 3dB, ideally 0 dB [7]. A low noise figure is crucial for various applications, such as signal processing [8], imaging [9] and communications [10].

There are two configurations for PSA; the first and more common is the degenerate signal setup, where a weak signal is spectrally placed between two pumps at equal detunings. Such amplifiers which were demonstrated in fibers [11] as well as a semiconductor waveguide [12] have one major drawback: a limited bandwidth. The second setup is based on a degenerate pump and two waves, idler and signal. Here, the frequencies where the gain takes place are determined by the dispersion since phase matching must be satisfied. If the pump propagates in the anomalous dispersion region, phase matching take place in the spectral vicinity of the pump resulting in broad band, highly effective gain. However, when the pump experiences normal dispersion, phase matching take place in two narrow band regions which are widely detuned from the pump [13]. The gain is less effective than in the broad band case but this configurations avails operation at wavelengths where no strong pump is easily available. The advantage of degenerate pump PSA over PIA was demonstrated in [14] where a sub 3dB noise figure was achieved. In a recent work, we have demonstrated a narrow band PS gain in the 2μm wavelength region using a pump whose wavelength is within the C band [15].

The main drawback of narrow band parametric gain is it bandwidth; for the 2μm amplifier, it was less than 0.5nm, and even for amplifiers with lower pump-gain detunings, the gain bandwidth is only several nm. In this work we introduce a modification to the PSA concept we have previously reported [15-16] where the degenerative pump is based on ps pulses with a variable spectral width. The use of a spectrally wide pump leads to two results: first, it broadens the gain bandwidth; a four to five fold increase compared to previous amplifiers (which were pumped by a quasi CW signal [16]) was demonstrated over a 300 nm spectral range. Second, changing the pump spectral width allows to control the amplifier bandwidth. Naturally, these amplifiers avail gain only during the short pulse duration. Hence they can serve mainly for signal processing and sampling applications rather than for communication where long data sequences need to be amplified.

The phase sensitive gain of the narrow band PSA manifests itself in a cyclical gain spectrum, in other words, the gain versus pump probe detuning (at the input) oscillates where negative gain (de-amplification) represents the regime where also the noise is squeezed. In the present PSA, the gain varies between -6 and 6 dB. The overall system gain was 28 dB.

## 2. Experimental Setup

The experimental setup for short pulsed pump PSA is schematically presented in Fig.1.

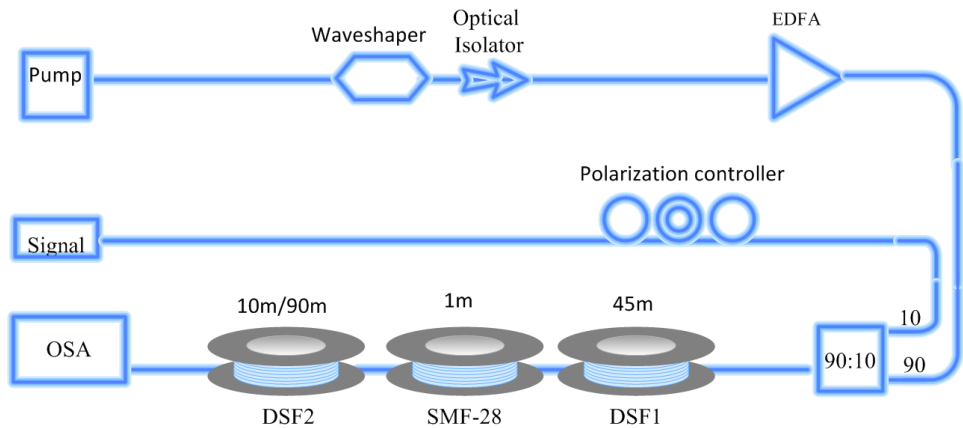

Figure 1. Experimental schematic

The pump is a pulsed laser emitting 14.6ps (with a spectral width of 0.34 nm) at a repetition rate of 20MHz and with a peak power at the laser output of up to 105 W. The pulses were amplified by an Erbium doped fiber amplifier (EDFA) before being combined with a CW signal. The amplifier consist of three fibers, the first is a 45 m long dispersion shifted fiber (DSF) whose zero dispersion wavelength (ZDW) is 1549 nm, and it's nonlinear coefficient $\gamma$ equals 2.4 $(W \cdot km)^{-1}$. An idler wave is generated in this fiber so that at it's output, there are three perfectly matched waves. These pass through the second fiber which is a short standard single mode dispersive fiber. The dispersive fiber modifies the relative phases of the three waves. As the detuning between the pump and signal at the input varies, the accumulated phase mismatch changes. When the three waves propagate in the third stage which is a DSF with properties similar to the stage one fiber, PSA takes place and cyclical gain is observed at the output. Maximum gain occurs at frequencies where phase matching conditions are satisfied while parametric de-amplification takes place at frequencies where destructive interference occurs.

## 3. Results

Fig.2. shows an output spectrum for a pump propagating at 1549nm which is at the edge of the anomalous dispersion regime. The high pump peak power leads to pump broadening, and a shift to shorter wavelength. Nevertheless, a clear PS gain is observed for over 100nm, with ASE variations of up to 13 dB.

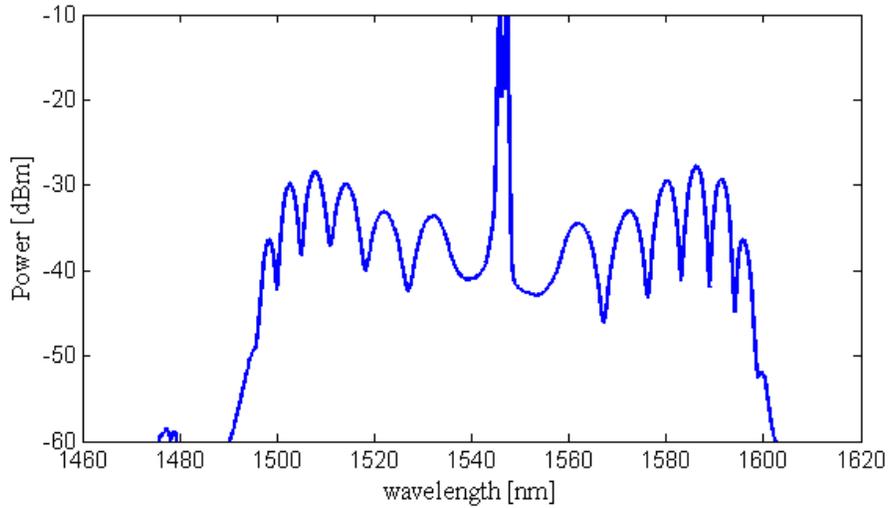

Figure 2. Broad band PSA output spectra

For a parametric amplifier with three waves: pump, signal and idler having angular frequencies $\omega_p$, $\omega_s$ and $\omega_i$, respectively the phase mismatch is $\Delta\beta=\beta_i+\beta_s-2\beta_p$, where $\beta_p$, $\beta_s$ and $\beta_i$ are the corresponding propagating constants. Gain is obtained at wavelengths where $-4P_0\gamma<\Delta\beta<0$, with $P_0$ being the pump peak power and $\gamma$ the fiber nonlinear coefficient. The edges of the gain region are at $\Delta\beta=0$ and $\Delta\beta=-4P_0\gamma$. For broad band amplification, the dependence on pump wavelength is not very important compared to the narrow band case and the wide bandwidth of the amplifier stems from high pump peak power.

For a pump propagating in the normal dispersion region, NB PSA is observed as seen in the measured results presented in Fig. 3 where the pump propagates at 1544 nm, and the varying parameter is the pump spectral width. The output spectra were measured for 3 pump spectral widths: 0.078nm (cyan), 0.15nm (red) and 0.3 nm (green). Also, the output spectrum for NB-PSA with a 4 ns pulsed pump is presented (blue), taken from [16]. The measurement for the 4 ns pump used a 1:10 output coupler and therefore the actual power level is 10dB higher, and matches the ps pulsed pump PSA case. Different spectral width of the pump lead to different pulse durations which change in turn the duty cycle and therefore the peak pulse power. Accurate determination of the pump peak power is complicated due to nonlinear broading and additional nonlinear processes that the pulse experience upon propagation. Assuming that the fiber nonlinear coefficient is constant, the peak ASE (and gain level) are determined by the peak power. In order to keep the peak power constant, the average pump power was set to 20.2 dBm for a 0.078 nm pulse spectral width, 19.4 dBm for 0.15 nm and 18.8 dBm for the 0.3 nm case.

Variations in the pump spectral width require separation between the effect on the gain of pump peak power and its spectral width. In order to achieve this, we measured the amplifier ASE for three pump spectral widths keeping the peak power constant. An enlarged view of the short (Fig.3(b)) and long (Fig.3(c)) wavelengths gain regimes reveals that indeed, the maximum ASE level and hence the peak powers, are similar (for the long wavelength) or even slightly higher for the 0.078 nm case (in the short wavelength), leading to the conclusion that the change in the amplifier bandwidth is caused by the widening of the pump spectrum. The overall bandwidth is 40 nm for a 0.3 nm wide pump, decreasing to 35 nm for the case of 0.15 nm, and shrinks to 30 nm for a pump spectral width of 0.078 nm. The bandwidth of the PSA with 4 ns pulses is also presented and is about 10 nm. The spectral width values are for the short wavelength region, but the picture is similar for the long wavelength.

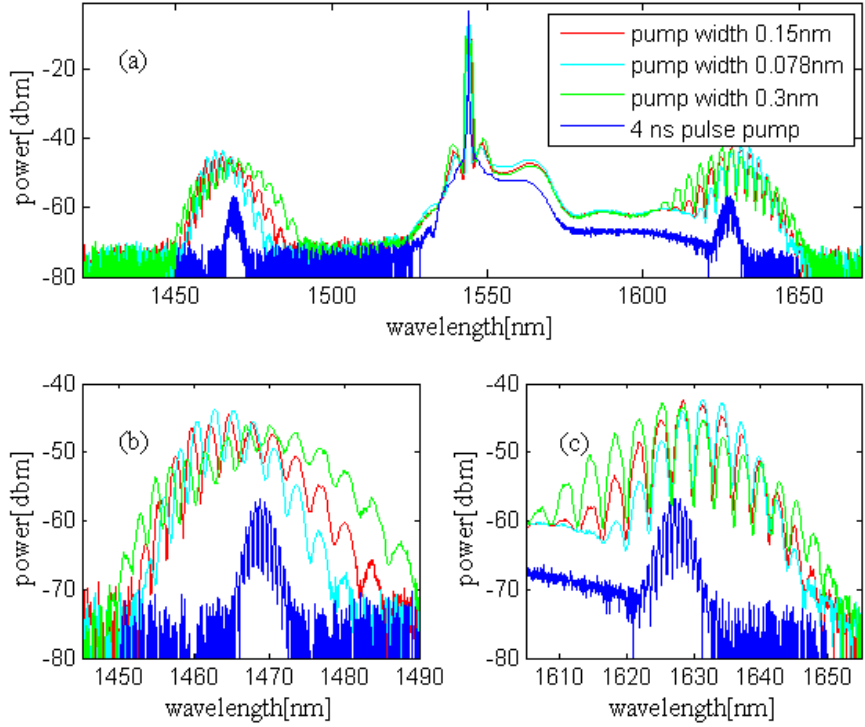

Figure 3. a) Output spectra of PSA for three pump spectral widths, and zoom on
b) short and c) long wavelength ASE regions

Next we characterized the gain by applying a tunable signal and sweeping it across the short wavelength ASE region. The pump propagated at 1545.5 nm, and had a spectral width of 0.078 nm. The pump induces two ASE zones centered at 1475 nm and 1625 nm. The output spectra of the amplifier with no applied signal is shown in Fig.4(a). Each of the ASE zones is 25nm wide and consists of 6 gain periods.

Sweeping a tunable signal across the short wavelength ASE region, induces an idler in the long wavelength gain region. The idler is measured at the input (Fig.4(b)) and the output (Fig.4(c)) of the third fiber, where PSA take place. At the input, we observe a conventional phase insensitive gain spectrum while at the output of the third fiber, a periodic pattern is observed due to phase sensitive process.

The gain pattern is calculated from the propagation equation for the three interacting waves [17]

$$\frac{dP_s}{dz} = -\alpha P_s + 2\gamma \left(P_p^2 P_s P_i\right)^{1/2} \sin(\theta), \qquad (1)$$

where $P_p$, $P_i$ and $P_s$ are the optical powers of the pump, idler and signal respectively, and $\alpha$ is the linear loss coefficient. The evolution of the idler is described by a similar equation. The angle $\theta$ represents the accumulated phase difference between the three fields at the output of the dispersive fiber. This difference is caused only by the linear dispersion since the dispersive fiber nonlinear coefficient is not sufficiently high to add a nonlinear phase contribution. Using the first two even order terms of the dispersion function $\beta(\omega)$, $\theta$ is, for a dispersive fiber of length L

$$\theta = (\beta_2(\omega-\omega_p)^2 + \beta_4(\omega-\omega_p)^4/12) \cdot L. \qquad (2)$$

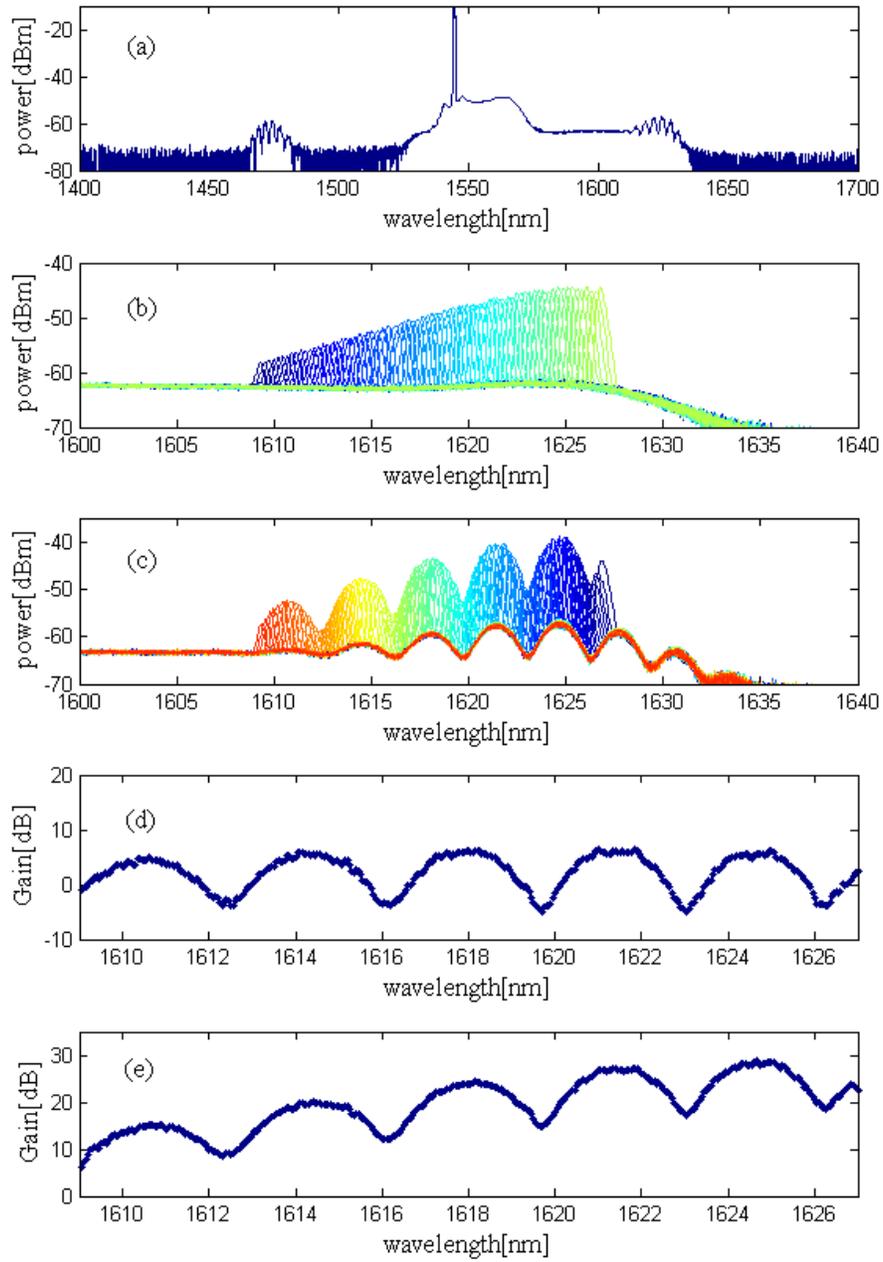

Figure 4. a) Output spectra of PSA with pump propagating at 1545.5nm and no signal applied. Generated idler at b) input and c) output of the third fiber. d) Calculated PSA gain, e) overall system on/off gain

For θ = π/2, the signal shows maximum gain, while for θ = -π/2, the rate of change of $P_s$ is negative which means that mode squeezing and hence de-amplification take place.

Since both input and output spectra were obtained, it is possible to directly derive the gain properties. The phase sensitive gain is presented at Fig.4(d), with gain variations between -6 and 6 dB. Each of the gain periods has a width of 3.5 nm and comprises six periods. Significant negative gain represents mode squeezing and is important for achieving a low noise figure. The overall system on/off gain is presented at Fig.4(e). It is determined from measurements of the idler conversion efficiency, taking also into account the duty cycle of the pump. The overall gain is up to 27dB, with a gain variance of 12 dB due to the phase sensitive amplification.

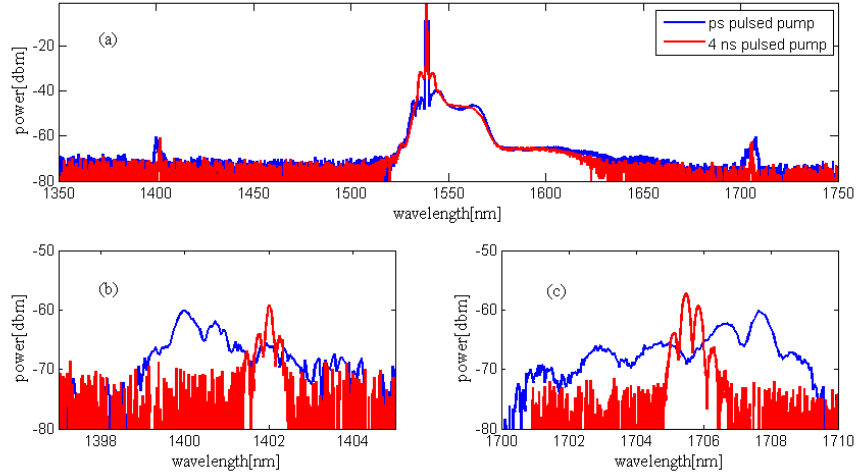

Figure 5. Comparison of short pulsed and quasi-CW pumped PSA a) output spectra with pump propagating at 1539nm, and zoom on b) short and c) long wavelength ASE regions

Fig.5 shows the amplifier performance at the edges of operation. The pump propagates at 1539 nm and has a spectral width of 0.05 nm. The two gain zones are centered at 1400nm and 1705nm (Fig.5(a)). The output spectrum is compared to that for a 4 ns pulsed pump. Fig. 5(b) and (c) show zoomed views of the ASE regions. Several gain periods are seen in both regions. The overall amplifier bandwidth is 4 times wider than for a PSA pumped by 4 ns pulses. On ps pulsed pump setup the dispersive fiber length was 1m, compared to 3 meters for 4 ns pulses.

### 4. Conclusions

To conclude, we have demonstrated a phase sensitive amplifier pumped by ps pulses. The amplifier operates in both broad and narrow band gain regime. The overall operation bandwidth of the amplifier is 100nm for the broad band regime, and up to 300 nm in the narrow band configuration. The use of short pump pulses leads to two results: first, a significant broadening of the gain spectrum. Second, and more important, is the ability to control the amplifier bandwidth by the pump spectral width. The ability to tune the gain bandwidth between 30 and 40 nm was demonstrated, by changing the pump spectral width from 0.078 to 0.3 nm. More control over the gain pattern, can be achieved by using more sophisticated amplitude and phase modifications of the pump pulse.